\documentclass[twocolumn,amsmath,amssymb,10pt,superscriptaddress,a4paper,letterpaper,fleqn]{revtex4-1}
\usepackage{amssymb}
\usepackage{epsfig}
\usepackage{graphicx}
\usepackage{dcolumn}
\usepackage{array}
\usepackage{bm}
\usepackage{fancyheadings}
\usepackage{longtable}
\usepackage{multirow}
\usepackage{float}
\usepackage{afterpage}  
\usepackage{color}
\usepackage{amsmath}

\setlongtables
\usepackage[breaklinks=true,linkbordercolor={1 1 1}]{hyperref}
\usepackage{hhline}

\begin{document}
\setcounter{page}{1}

\title{
\qquad \\ \qquad \\ \qquad \\  \qquad \\  \qquad \\ \qquad \\ 
Systematics of Evaluated  Half-Lives of Double-Beta Decay}

\author{Boris Pritychenko}
\email[Corresponding author, electronic address:\\ ]{pritychenko@bnl.gov}
\affiliation{National Nuclear Data Center, Brookhaven National Laboratory, Upton, NY 11973-5000, USA}

\date{\today} 

\begin{abstract}
{
A new  evaluation of 2$\beta$-decay half lives and their systematics is presented. These data extend the previous evaluation and include the analysis of all recent measurements.  The nuclear matrix elements for 2$\beta$-decay transitions in 12 nuclei have been extracted. The recommended values are compared with the large-scale shell-model, QRPA calculations, and experimental data. 
A T$^{2\nu}_{1/2 }$ $\sim$ 1/E$^{8}$ systematic trend has been observed for recommended $^{128,130}$Te  values. This trend indicates similarities for nuclear matrix elements in Te nuclei and was predicted for 2$\beta$(2$\nu$)-decay mode.  The complete list of  results is available online at http://www.nndc.bnl.gov/bbdecay/.}
\end{abstract}
\maketitle


\lhead{ND 2013 Article $\dots$}
\chead{NUCLEAR DATA SHEETS}
\rhead{Boris Pritychenko}
\lfoot{}
\rfoot{}
\renewcommand{\footrulewidth}{0.4pt}


\section{ INTRODUCTION}
Double-beta decay was originally proposed by Goeppert-Mayer in 1935 \cite{35Goe} as a nuclear disintegration with simultaneous emission of  two electrons and two neutrinos
\begin{equation}
\label{myeq.1a}  
(Z,A) \rightarrow (Z+2,A) + 2 e^{-} + 2\bar{\nu}_{e}.
\end{equation}

 There are several double-beta decay processes: $2 \beta^{-}$,  $2 \beta^{+}$, $\epsilon$ $\beta^{+}$, 2$\epsilon$ and  decay modes: 
 two-neutrino (2$\nu$), neutrinoless (0$\nu$) and Majoron emission ($ \chi^{0}$)
 \begin{equation}
\label{myeq.1b}  
(Z,A) \rightarrow (Z\pm2,A) + (2 e^{\pm}) + (2\bar{\nu}_{e},\  2{\nu}_{e}\ or\  \chi^{0}).
\end{equation}
 The 2$\nu$-mode is not prohibited by any conservation law and definitely occurs as a second-order process compared with regular $\beta$-decay \cite{92Boe}. The 0$\nu$-mode differs from the 2$\nu$-mode in that only electrons  are emitted during the decay. This normally requires that the lepton number is not conserved and the neutrino should contain a small fraction of massive particles equal to its anti-particles (Majorana neutrino). Obviously, observation of $2 \beta$(0$\nu$)-decay will have significant implications for particle physics and fundamental symmetries, while observation of $2 \beta$(2$\nu$)-decay will provide information on nuclear structure physics that can be used in 0$\nu$-mode calculations.  

Historically, the search for double-beta decay has been a very hot topic in nuclear physics \cite{11Tre,02Tre}. Nuclear physicists and chemists employed a variety of direct (nuclear radiation detection) and geochemical methods.  In recent years  claims have been made for the observation of the $0\nu$-decay mode in $^{76}$Ge \cite{01Kl}. These rather controversial results were widely scrutinized and often rejected by the nuclear physics community \cite{02Aa,05Ba}. However, the 2$\nu$-decay mode has  definitely been  observed in many isotopes. Table \ref{table0} provides a brief review of 2$\beta$-decay observations. Because of the extremely low probability for double beta decay it was  first detected by analyzing the chemical composition of rock samples and later verified by more accurate direct  detection methods.

\begin{table*}[!htb]
\centering
\caption{Brief history of experimental observations  of 2$\beta$-decay and  reported {\it T}$_{1/2}$(2$\beta$) values. Results are shown for direct detection   and geochemical methods.}
\begin{tabular}{c|c|c|c|c|c|c|c}
\hline
\hline
\multirow{2}{*}{Parent nuclide}  &  \multirow{2}{*}{Process} & \multirow{2}{*}{Transition} &  \multicolumn{2}{c}{Discovery year}  & \multicolumn{3}{|c}{Originally  reported {\it T}$_{1/2}$($\beta$$\beta$), (y)} \\ 
\hhline{~~~-----}
 &  & & Direct & Geochemical & 2$\nu$, direct & (2+0)$\nu$, direct  & (2+0)$\nu$, geochemical \\ \hline
$^{48}$Ca & $2 \beta^{-}$ & 0$^{+}$ $\rightarrow$ 0$^{+}$ & 1996 & & 4.3x10$^{19}$ \cite{48Ca}  & &  \\ 
$^{76}$Ge & $2 \beta^{-}$  & 0$^{+}$ $\rightarrow$ 0$^{+}$ & 1990  &   & 9.0x10$^{20}$ \cite{76Ge} &  &  \\ 
$^{82}$Se & $2 \beta^{-}$  & 0$^{+}$ $\rightarrow$ 0$^{+}$ &  1987  & 1969 & 1.1x10$^{20}$ \cite{82Se2}  & & 1.4x10$^{20}$ \cite{82Se1} \\ 		
$^{96}$Zr & $2 \beta^{-}$  & 0$^{+}$ $\rightarrow$ 0$^{+}$  &  1999 & 1993 &  2.1x10$^{19}$ \cite{96Zr2} & & 3.9x10$^{19}$ \cite{96Zr1} \\
$^{100}$Mo & $2 \beta^{-}$  & 0$^{+}$ $\rightarrow$ 0$^{+}$   &  1990 & & 3.3x10$^{18}$ \cite{100Mo} & &  \\ 	
$^{100}$Mo & $2 \beta^{-}$  & 0$^{+}$ $\rightarrow$ 0$^{+}_{1}$   &  1995 & &  9.5x10$^{18}$ \cite{100Mo1} &   & \\ 
$^{100}$Mo & $2 \beta^{-}$  & 0$^{+}$ $\rightarrow$ 0$^{+}_{1}$   &  1995 & &  & 6.1x10$^{20}$  \cite{100Mo2} & \\
$^{116}$Cd & $2 \beta^{-}$  & 0$^{+}$ $\rightarrow$ 0$^{+}$   & 1995 &  & 2.7x10$^{19}$ \cite{116Cd} &  & \\ 
$^{128}$Te & $2 \beta^{-}$  & 0$^{+}$ $\rightarrow$ 0$^{+}$   &   & 1975 & & & 1.5x10$^{24}$ \cite{128Te}  \\
$^{130}$Te & $2 \beta^{-}$  & 0$^{+}$ $\rightarrow$ 0$^{+}$   & 2003 & 1966 &  6.1x10$^{20}$ \cite{130Te2} & & 8.2x10$^{20}$ \cite{130Te1}  \\ 	
$^{136}$Xe & $2 \beta^{-}$  & 0$^{+}$ $\rightarrow$ 0$^{+}$   &  2011 & & 5.5x10$^{21}$ \cite{136Xe} &   &  \\	
$^{130}$Ba & 2$\epsilon$ & 0$^{+}$ $\rightarrow$ 0$^{+}$   &    & 2001 & & & 2.2x10$^{21}$ \cite{130Ba} \\
$^{150}$Nd & $2 \beta^{-}$  & 0$^{+}$ $\rightarrow$ 0$^{+}$  & 1993 & & 1.7x10$^{19}$ \cite{150Nd} &  & \\ 
$^{150}$Nd & $2 \beta^{-}$  & 0$^{+}$ $\rightarrow$ 0$^{+}_{1}$   &  2004 & &  & 1.4x10$^{20}$ \cite{150Nd1} & \\
$^{238}$U  & $2 \beta^{-}$  & 0$^{+}$ $\rightarrow$ 0$^{+}$   & & 1991 & & & 2.0x10$^{21}$ \cite{238U} \\
\hline
\hline
\end{tabular}
\label{table0}
\end{table*} 

Experimental evidence and theoretical calculations indicate that the probability for the 2$\nu$-mode is much higher than that for the 0$\nu$-mode.  In fact, $^{76}$Ge $2 \beta$-decay measurements   have demonstrated that the decay rate for $2 \beta$(2$\nu$)-decay is at least four  orders of magnitude higher than that of $2 \beta$(0$\nu$).  Therefore, we will concentrate on the experimentally observed 2$\nu$-mode only.

\section{COMPILATION AND EVALUATION OF EXPERIMENTAL DATA}
\begin{figure}[!htb]
\fbox{
\includegraphics[width=0.9\columnwidth]{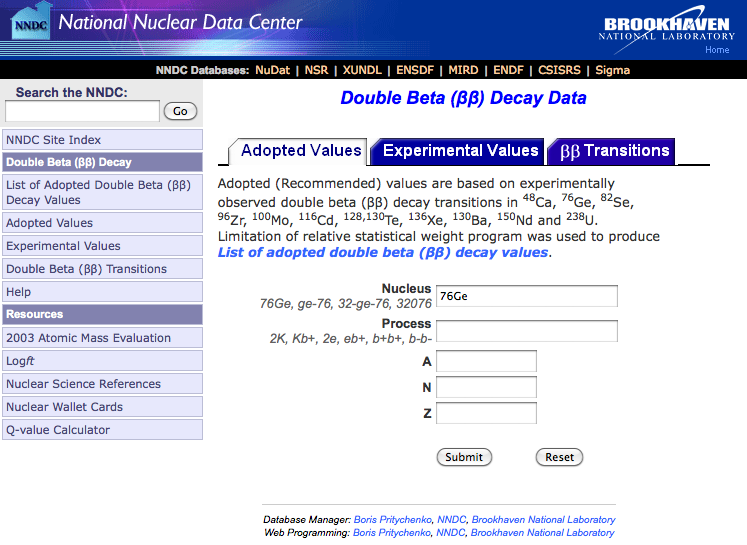}
}
\caption{The NNDC 2$\beta$-decay data  website   http:// www.nndc.bnl.gov/bbdecay/ \cite{06Prt,06Pri}.} 
\label{fig1}
\end{figure}

Double-beta decay is an important nuclear physics process and experimental results in this field have been compiled by several groups \cite{02Tre,06Prt,10Nak}.  Fig. \ref{fig1}  shows the online compilation and evaluation conducted at the  National Nuclear Data Center    since 2006 \cite{06Prt,06Pri}. Observed 2$\beta$-decay  data 
for  isotopes of interest are shown in Table \ref{table0}. While Table   \ref{table0} lists only a single paper per nuclide for direct and geochemical discovery methods,  a complete compilation is available from the NNDC website   http:// www.nndc.bnl.gov/bbdecay/. This compilation of experimental results includes results of previous \cite{02Tre} and recent work obtained by searching the Nuclear Science References database  \cite{10Pri,13Pri}.  It  was used to produce evaluated or recommended values.

\begin{table*}[!htb]
\centering
\caption{Recommended  {\it T}$_{1/2}$(2$\beta$) and complimentary parameter values. }
\begin{tabular}{c|c|c|c|c|c|c}
\hline
\hline
Parent nuclide  &  Process & Transition & Q-value (keV) & $\beta_2$ & {\it T}$_{1/2}^{2 \nu}$(y) &  {\it T}$_{1/2}^{2 \nu + 0 \nu}$(y)  \\
\hline

$^{48}$Ca & $2 \beta^{-}$ & 0$^{+}$ $\rightarrow$ 0$^{+}$ & 4267.0 & 0.2575(56) & (4.39$\pm$0.58)x10$^{19}$ & \\ 		
$^{76}$Ge & $2 \beta^{-}$  & 0$^{+}$ $\rightarrow$ 0$^{+}$ & 2039.06  & 0.3133(+55-20) & (1.43$\pm$0.53)x10$^{21}$ &  \\ 		
$^{82}$Se & $2 \beta^{-}$  & 0$^{+}$ $\rightarrow$ 0$^{+}$ &  2996.4  & 0.2031(+30-28) & (9.19$\pm$0.76)x10$^{19}$ & \\ 		
$^{96}$Zr & $2 \beta^{-}$  & 0$^{+}$ $\rightarrow$ 0$^{+}$ & 3349.0  &  0.1525(27) & (2.16$\pm$0.26)x10$^{19}$ &  \\ 		
$^{100}$Mo & $2 \beta^{-}$  & 0$^{+}$ $\rightarrow$ 0$^{+}$ &  3034.37  &  0.21539(90) & (6.98$\pm$0.44)x10$^{18}$ &  \\ 		
$^{100}$Mo & $2 \beta^{-}$  & 0$^{+}$ $\rightarrow$ 0$^{+}_{1}$ & 2339.3  &  & (5.70$\pm$1.36)x10$^{20}$ & \\ 		
$^{100}$Mo & $2 \beta^{-}$  & 0$^{+}$ $\rightarrow$ 0$^{+}_{1}$ & 2339.3  & & & (6.12$\pm$0.20)x10$^{20}$ \\
$^{116}$Cd & $2 \beta^{-}$  & 0$^{+}$ $\rightarrow$ 0$^{+}$ & 2813.44  & 0.1083(18) & (2.89$\pm$0.25)x10$^{19}$ &  \\ 		
$^{128}$Te & $2 \beta^{-}$  & 0$^{+}$ $\rightarrow$ 0$^{+}$ & 866.5  &  0.1862(37) & & (3.49$\pm$1.99)x10$^{24}$ \\
$^{130}$Te & $2 \beta^{-}$  & 0$^{+}$ $\rightarrow$ 0$^{+}$ & 2527.51  & 0.1630(+38-28) & (7.14$\pm$1.04)x10$^{20}$ &  \\ 	
$^{136}$Xe & $2 \beta^{-}$  & 0$^{+}$ $\rightarrow$ 0$^{+}$ & 2457.99  &  0.1262(17) & (2.34$\pm$0.13)x10$^{21}$ &    \\	
$^{130}$Ba & 2$\epsilon$ & 0$^{+}$ $\rightarrow$ 0$^{+}$ & 2620.1  &   0.1630(+38-28) & & (1.40$\pm$0.80)x10$^{21}$ \\
$^{150}$Nd & $2 \beta^{-}$  & 0$^{+}$ $\rightarrow$ 0$^{+}$ &  3371.38  & & (8.37$\pm$0.45)x10$^{18}$ & \\ 		
$^{150}$Nd & $2 \beta^{-}$  & 0$^{+}$ $\rightarrow$ 0$^{+}_{1}$ &  2696.0  &  & & (1.33$\pm$0.40)x10$^{20}$ \\
$^{238}$U  & $2 \beta^{-}$  & 0$^{+}$ $\rightarrow$ 0$^{+}$ & 1144.2  & & & (2.00$\pm$0.60)x10$^{21}$ \\
\hline
\hline
\end{tabular}
\label{table1}
\end{table*} 
Table \ref{table1} shows  the latest recommended values  which were deduced in the accordance with the US Nuclear Data Program guidelines \cite{98Lwe,06Bur}. All final results from independent observations were included in the evaluation process. These evaluated half-lives represent the best  values currently available; further measurements will result in the addition of new and improved values. Table \ref{table1} also includes recent data on decay Q-values  \cite{Wa12,12Pri} and quadrupole deformation parameters which are used in the next section.

\section{ANALYSIS OF RECOMMENDED VALUES}

To separate nuclear structure effects from the kinematics, the nuclear matrix elements for $\beta$$\beta$(2$\nu$)-decay were extracted from the present evaluation of half-lives.  $T_{1/2}^{2\nu}$ values are often described as follows \cite{92Boe}
 \begin{equation}
\label{myeq.Half}
 \frac{1}{T_{1/2}^{2\nu} (0^{+} \rightarrow 0^{+})} = G^{2 \nu} (E,Z)  |M^{2 \nu}_{GT} - \frac{g^{2}_{V}}{g^{2}_{A}} M^{2 \nu}_{F}|^{2}, 
 \end{equation}
 where the function  $G^{2 \nu} (E,Z)$ results from lepton phase space integration and contains all the relevant constants. 
Table \ref{table2} shows the effective nuclear matrix elements  (M$_{eff}^{2\nu}$) for  $\beta$$\beta$(2$\nu$)-decay based on the latest phase factor calculation from the Yale group  \cite{12Kot}.

\begin{table*}
\centering
\caption{Effective nuclear matrix elements  (M$_{eff}^{2\nu}$) for  2$\beta$(2$\nu$)-decay from the present work, ITEP evaluation, large-scale shell-model and QRPA calculations.}
\begin{tabular}{c|c|c|c|c|c|c}
\hline
\hline
Parent nuclide  &  Process & Transition & Present work &  Yale \& ITEP \cite{12Kot,10Bar} &  Shell model \cite{12Cau}   &  QRPA \cite{12Rad} \\
\hline
$^{48}$Ca & $2 \beta^{-}$ & 0$^{+}$ $\rightarrow$ 0$^{+}$ & 0.0383$\pm$0.0025 & 0.038$\pm$0.003 & 0.0389,0.0397,0.0538 & 0.0373 \\
$^{76}$Ge & $2 \beta^{-}$  & 0$^{+}$ $\rightarrow$ 0$^{+}$ & 0.120$\pm$0.021 & 0.118$\pm$0.005 & 0.0961 & 0.147 \\
$^{82}$Se & $2 \beta^{-}$  & 0$^{+}$ $\rightarrow$ 0$^{+}$ & 0.0826$\pm$0.0034 & 0.083$\pm$0.004 & 0.104 & 0.0687 \\
$^{96}$Zr & $2 \beta^{-}$  & 0$^{+}$ $\rightarrow$ 0$^{+}$ & 0.0824$\pm$0.0050 & 0.080$\pm$0.004 & & 0.0952 \\
$^{100}$Mo & $2 \beta^{-}$  & 0$^{+}$ $\rightarrow$ 0$^{+}$ & 0.208$\pm$0.007 & 0.206$\pm$0.007 &  & 0.183 \\
$^{100}$Mo & $2 \beta^{-}$  & 0$^{+}$ $\rightarrow$ 0$^{+}_{1}$ & 0.170$\pm$0.020 & 0.167$\pm$0.011  &  & \\
$^{116}$Cd & $2 \beta^{-}$  & 0$^{+}$ $\rightarrow$ 0$^{+}$ & 0.112$\pm$0.005 & 0.114$\pm$0.005 &  & 0.132 \\
$^{128}$Te & $2 \beta^{-}$  & 0$^{+}$ $\rightarrow$ 0$^{+}$ & 0.0326$\pm$0.0093 & 0.044$\pm$0.006 & 0.0489,0.0306 & 0.0464 \\
$^{130}$Te & $2 \beta^{-}$  & 0$^{+}$ $\rightarrow$ 0$^{+}$ & 0.0303$\pm$0.0022 & 0.031$\pm$0.004 & 0.0356,0.0224 & 0.019 \\
$^{136}$Xe & $2 \beta^{-}$  & 0$^{+}$ $\rightarrow$ 0$^{+}$ & 0.0173$\pm$0.0005 &   & 0.0207 & \\
$^{130}$Ba & 2$\epsilon$ & 0$^{+}$ $\rightarrow$ 0$^{+}$ & 0.218$\pm$0.062 & 0.174$\pm$0.017 & & \\
$^{150}$Nd & $2 \beta^{-}$  & 0$^{+}$ $\rightarrow$ 0$^{+}$ & 0.0572$\pm$0.0015 & 0.058$\pm$0.004 & & 0.0348 \\
$^{150}$Nd & $2 \beta^{-}$  & 0$^{+}$ $\rightarrow$ 0$^{+}_{1}$ & 0.0417$\pm$0.0063 & 0.042$\pm$0.006 & & \\
$^{238}$U  & $2 \beta^{-}$  & 0$^{+}$ $\rightarrow$ 0$^{+}$ & 0.185$\pm$0.028 & 0.19$\pm$0.04 &  & \\
\hline
\hline
\end{tabular}
\label{table2}
\end{table*} 

The present results can be compared with the Yale University re-evaluation of the ITEP  data \cite{10Bar}.  
The major differences between the present work  and ITEP  include  $^{128}$Te and $^{136}$Xe evaluated half-lives and the general evaluation philosophy.
\begin{itemize}
\item $^{128}$Te: The ITEP evaluation rejects one geochemical result \cite{128Te1} as a possible indication of changing  weak interaction constants over the last billion years and adopts the second one   with  corrections \cite{128Te2,10Bar}. The present evaluation is based on the final published results of five measurements without any corrections.
\item $^{136}$Xe: These data became available a year later than publication of the ITEP evaluation. The NNDC half-life value is based on the results from three independent groups \cite{136Xe,136Xe1,136Xe2}.
\item The ITEP evaluation treats all (2+0)$\nu$ observations as pure 2$\nu$-decay mode results and includes many other assumptions that allow deduction of  the precise  values of nuclear matrix elements based on very limited statistics. However, the present evaluation clearly indicates large uncertainties for nuclear matrix elements.
\item Finally, this work uses the latest values of the phase factors \cite{12Kot}, while ITEP is based on rather outdated values \cite{92Boe,98Suh}.
\end{itemize}

The evaluated nuclear matrix elements  can be compared with recent theoretical calculations of $M^{2 \nu}_{GT}$ \cite{12Cau,12Rad} using the following equation \cite{12Kot}
\begin{equation}
\label{myeq.2b}  
|M_{eff}^{2\nu}|= g^{2}_{A} \times |(m_{e}c^2) M^{2 \nu}_{GT}|, 
\end{equation}
where $g^{2}_{A}$=1.273$^{2}$ and $m_{e}c^2$=0.511 MeV. Analysis  of the data in Table \ref{table2} indicates reasonably good agreement between theoretical and experimental values of the nuclear matrix elements. Several deviations are due to problems with the calculation of nuclear matrix elements for very weak decays \cite{12Hor} because  accurate values of  the Gamow-Teller strength functions are often missing.

To gain a better understanding of decay half-lives, we will analyze the half-life values of $^{128,130}$Te in more detail. Both tellurium isotopes have the same charge and a similar shell structure and deformation,  but the $2 \beta^{-}$-transition energies are different.  It is natural to assume that the difference between tellurium half-lives is due to transition energies \cite{68Pon}. In fact, in the present evaluation,  central values for $T_{1/2}^{2\nu}$ are consistent with the following ratio
\begin{equation}
\label{myeq.2c}  
\frac{T_{1/2}^{2\nu} (^{128}Te)}{T_{1/2}^{2\nu} (^{130}Te)} \approx 4.9 \times 10^{3} \sim (\frac{E_{^{130}Te}}{E_{^{128}Te}})^{7.9}.
\end{equation}
From this equation we deduce the following systematic trend
\begin{equation}
\label{myeq.2d}  
T_{1/2}^{2\nu} (0^{+} \rightarrow 0^{+}) \sim \frac{1}{E^8}.
\end{equation} 

This conclusion agrees  well with the theoretical calculation of Primakoff and Rosen \cite{69Pri} who predicted that for 2$\beta$(2$\nu$) decay, the phase space available to the four emitted leptons is roughly proportional to the 8$^{th}$ through 11$^{th}$ power of energy release. It is worth noting that in many direct detection experiments the discovery was based on the observation of the total energy deposition, and authors often could not separate a two-electron event from the single-electron tracks  \cite{06Prt}. Consequently, the observed relation between experimental half-lives and transition energies provides an additional observable quantity for double-beta decay processes.

Additional analysis of $^{96}$Zr, $^{100}$Mo,  $^{130}$Te, and $^{136}$Xe decay rates provides complimentary experimental evidence that deformation strongly affects the half-life values. For example, it is easy to see that the lower recommended value for   $^{100}$Mo  vs. $^{96}$Zr cannot be explained by transition energy or electric charge   contributions; a similar situation exists for   $^{130}$Te vs. $^{136}$Xe. These examples show that well-known experimental values of quadrupole deformation parameters could help to understand the relations between recommended half-lives when appropriate Gamow-Teller  strength functions are not available from charge-exchange reactions \cite{12Cau} or are extremely  difficult to measure \cite{13Ru}.

\section{ CONCLUSIONS}
Double-beta decay is a very rare nuclear physics process that is often used to test theoretical  model predictions for elementary particle and nuclear structure physics. The present work contains the latest evaluation of the experimental half-lives and nuclear matrix elements. The nuclear matrix elements strongly rely on phase factor  calculations that can vary \cite{12Kot,98Suh,92Boe}. This implies the importance of experimental {\it T}$_{1/2}^{2 \nu}$  compilation and evaluation as a primary model-independent quantity. 

The compilation and analysis of experimental papers \cite{06Prt} indicates a strong interest in double-beta decay over the past 75 years. Several new measurements have been performed recently and many others are under way. This is why online compilation and 2$\beta$-decay  data dissemination play essential roles.  Continuing research and observation of additional decay properties will help to clarify the situation by comparing the observables with theoretical predictions. The  $^{128,130}$Te half-lives and their systematic trend could play a crucial role in our understanding of the interplay of phase factors and 2$\beta$(2$\nu$)-decay nuclear matrix elements,   which will eventually lead to an overall improvement of theoretical models and better interpretation of experimental results.

Future work on the double beta-decay horizontal evaluation and compilation will be conducted in collaboration with KINR, Ukrainian Academy of Sciences.


{\it Acknowledgments:} We are  indebted to M. Herman (BNL) for support of this project and V. Tretyak (KINR) for useful suggestions. 
We are also grateful to V. Unferth (Viterbo University) and M. Blennau (BNL) for their help with the manuscript. This work was funded by the Office of
Nuclear Physics, Office of Science of the US Department of Energy, under Contract No. DE-AC02-98CH10886 with
Brookhaven Science Associates, LC.

\end{document}